\documentclass{appolb}
\usepackage{epsfig}
\usepackage{axodraw}

\begin{document}
\pagestyle{plain}
\newcount\eLiNe\eLiNe=\inputlineno\advance\eLiNe by -1
\title{{Neutrino Physics and CP violation
\thanks{Invited lectures given by G. C. Branco at 
the 47th Cracow School of Theoretical 
Physics, Zakopane, Poland}
}
\author{G. C. Branco and M. N. Rebelo
\address{Departamento de F{\'\i}sica and Centro  de F{\'\i}sica
Te{\'o}rica de Part\'\i}culas (CFTP),
Instituto Superior T\'{e}cnico (IST), Av. Rovisco Pais, 1049-001
Lisboa, Portugal}}
\maketitle

\begin{abstract}
We review some aspects of neutrino physics and CP violation
both in the quark and lepton sectors.
\end{abstract}

\section{Introduction}

In these lectures, we cover topics related to neutrino physics and CP
violation. The treatment of these topics is not
extensive. For a thorough treatment of the above topics,
the reader is advised to consult the excellent books and review articles 
which exist in the literature covering neutrino physics and 
CP violation. Some books are cited in what follows. A list of a few 
review articles can be found in Ref.~\cite{reviews}. These lectures
are organised as follows. In the next section we describe some of the 
minimal extensions of the SM which can incorporate nonvanishing
neutrino masses, with special emphasis on the seesaw
mechanism  \cite{seesaw}. In section three we cover CP violation 
both in the quark and lepton sector. In the lepton sector, 
we discuss CP violation both at low and high energies in the minimal
seesaw mechanism. In the last subsection we briefly describe the
generation of the baryon asymmetry of the Universe through leptogenesis.

\section{Minimal extensions of the SM  incorporating neutrino masses}
In the leptonic sector of the SM the fermionic field content is:
\begin{equation}
L_{Li} = \left(\begin{array}{c}
{\nu^0} \\
l^0  \end{array} \right)_{Li}, \quad  l^0_{Ri}, \quad  (i=1,2,3)
\end{equation}
where $L_{Li}$ denote the lefthanded leptonic doublets, containing 
neutrinos and charged leptons. The righthanded components of the
charged leptons, $l^0_{Ri}$, are SU(2) singlets.
No righthanded components for the neutrino fields are introduced
in the SM. 

The charged leptons acquire mass through Yukawa terms 
of the form:
\begin{equation}
{\cal L}_Y = f_{ij}^e {\overline L_{Li}} l^0_{Rj}\phi + hc .,
\end{equation}
With $\phi$ a scalar Higgs doublet:
\begin{equation}
\phi =  \left(\begin{array}{c} \phi^+ \\ \phi^0 
\end{array}\right) \label{higgs}
\end{equation}
Due to the absence of righthanded singlet fields  ${\nu^0}_{Ri}$, 
it is not possible to
have a Dirac mass term for the neutrinos.

In general, Dirac mass terms are of the form 
\begin{equation}
-{\cal L}^m_D = m_D \overline\psi \psi = m_D (\overline\psi_L \psi_R +
\overline\psi_R \psi_L )
\end{equation}
and are invariant under $U(1)$ transformations, i.e., they
conserve any charge carried by $\psi$ associated to a $U(1)$  
symmetry (e.g., electrical charge, 
lepton number, etc). Upon spontaneous gauge symmetry breaking (SSB)
the neutral component of $\phi$ acquires a vaccum expectation value 
\begin{equation}
< \phi^0 > = \frac{v}{\sqrt2} 
\end{equation}
As a result, Dirac mass terms for the charged leptons are generated from
the Yukawa couplings, given by:
\begin{equation}
{\cal L}_{m_l} = \frac{v}{\sqrt2} \ f_{ij}^e \ {\overline l^0_{Li}} 
\ l^0_{Rj}
+ h.c. \ \equiv \ {m_l}_{ij}  \ {\overline l^0_{Li}} \ l^0_{Rj} + h.c.
\label{tric}
\end{equation}

Neutrinos have the very special feature of being the only known
fermions which have zero electrical charge. As a result, neutrinos
can have Majorana mass terms, which are of the form
\begin{equation}
-{\cal L}^m_M = \frac{1}{2} \left[ \psi_L^T C m_L \psi_L + hc \right] 
\end{equation} 
where $C$ is the charge-conjugation matrix defined by:
\begin{equation}
C^{-1} \gamma_\mu C = - \gamma^T_{\mu}
\end{equation} 
with $\gamma_\mu $ denoting the Dirac matrices.

With the fermionic content of the Standard Model this would 
correspond to terms of the form  $\nu_{Li}^{0T} C \nu_{Lj}^0 $.
However, in the SM these terms cannot be introduced at the
Lagrangian level, because they are not gauge invariant. Also,
due to exact B-L conservation, they cannot be generated neither
radiatively in higher orders, nor nonperturbatively. As a
result, neutrinos are {\it strictly massless} in the SM. 

The charged 
lepton mass matrix given in Eq.~(\ref{tric}) can be diagonalised 
through the biunitary transformation:
\begin{equation}
U^{\dagger}_{l L} \ m_l \ U_{l R} = \mbox{diag} 
\ (m_e, \ m_\mu, \ m_\tau ) 
\end{equation} 
After this transformation, the matrix $U_{l L}$ would appear in the charged
leptonic interactions:
\begin{equation}
{\cal L}_W =  \frac{v}{\sqrt 2}\  {\overline l_{L} \ \gamma_{\mu}}\ 
U^{\dagger}_{l L} \ \nu_L \ W^{+\mu}  + h. c.
\end{equation} 
However, since in the SM neutrinos are strictly massless, the matrix
$U^{\dagger}_{l L}$ can always be eliminated through a redefinition of
the neutrino fields. Therefore in the SM there is {\it no leptonic mixing},
which leads to separate conservation of the flavour lepton numbers
$L_e$, $L_\mu$, $L_\tau$. Consequently the recent observation of neutrino
oscillations provides clear evidence for {\it physics beyond the SM}.

\subsection{Generating neutrino masses through extensions of the scalar
sector}
There are various ways of generating neutrino masses through
extensions of the SM involving the scalar sector \cite{Mohapatra:1998rq}.
Since $\nu_{L}$ are part of a doublet, one of the simplest ways
of generating such mass terms is by extending the SM
through the introduction of a scalar Higgs triplet \cite{Cheng:1980qt}
$\vec H$ which would allow for the following renormalizable Yukawa term.
\begin{equation}
-{\cal L}^H_Y = f_{ij}  L^T_{Li} C (i \tau_2)(\vec \tau \cdot \vec H) L_{Lj}
+h.c.
\end{equation}
still conserving lepton number, where
\begin{equation}
\vec \tau \cdot \vec H = \left(\begin{array}{cc}
H^+ & \sqrt 2 H^{++} \\
\sqrt 2 H^0  & -H^+ \end{array}\right) \label{lota}
\end{equation}
When $\vec H$ develops a vacuum expectation value (vev) lepton number is
violated and Majorana mass terms for the neutrinos are generated.
Notice that $[f_{ij}]$ is a symmetric matrix due to anticommutation
of the fermion fields, the antisymmetric property of the charge
conjugation matrix and the symmetric character of $(i \tau_2) \vec \tau $.
Note that even in the context of the SM, it
would be possible to construct a composite triplet Higgs operator
out of two Higgs doublets. Of course, such a term 
$(L^T_{L} C i \tau_2\vec \tau  L_{L})(\phi ^T i \tau_2 \vec \tau \phi)$
has dimension five and would be non renormalizable. However,
it cannot be effectively generated in the Standard Model since it violates
$B-L$ which is an accidental exact symmetry of the Standard Model.

An alternative simple way of generating Majorana mass terms for lefthanded
neutrinos is, for instance, the introduction of a singly charged scalar 
singlet $h^+$ as proposed by Zee \cite{Zee:1980ai}
allowing for a Yukawa coupling 
of the form:
\begin{equation}
-{\cal L}^H_Y = f_{ij}  L^T_{Li} C (i \tau_2) L_{Lj} h^+ +h.c.
\label{2a}
\end{equation}
in this case $[f_{ij}]$ must be an antisymmetric matrix,
since $i \tau_2 \equiv \varepsilon $ is antisymmetric. 
This coupling by itself does not violate $B-L$, since one has the freedom
to assign $B-L$ quantum number $(-2)$ to the field $h^+$.
In order to generate neutrino masses, one needs at least two Higgs doublets
and a cubic coupling of the form 
\begin{equation}
M_{\alpha \beta} \varepsilon _{ij} \phi^i_{\alpha} \phi^j_{\beta} h^-
\label{2b}
\end{equation}
where the indices $\alpha$, $\beta$ distinguish between the Higgs doublets
and $i, j$ are SU(2) indices. The coupling  $M_{\alpha \beta}$
has dimension of mass and is antisymmetric. The simultaneous 
presence of the two couplings (\ref{2a}) and (\ref{2b}) in the theory
violates explicitly $B-L$ by two units and leads to finite 
and calculable one loop contributions to neutrino masses.

These are just two of the simplest examples considered 
in the literature, where neutrino masses are generated via
extensions of the scalar sector. 

\subsection{Generating neutrino masses through the introduction
of righthanded neutrinos}
In these lectures we are mainly concerned with 
extensions of the Standard Model where only $SU(2) \times U(1)$
singlet righthanded neutrinos are added to its spectrum.
Indeed one may view the simple addition of right-
handed
neutrino components to the SM as the most straightforward way of
incorporating neutrino masses. In this case the number of 
fermionic degrees of freedom for neutrinos equals those of all other fermions
in the theory provided that three righthanded neutrinos are introduced.
It is well known 
that such an extension of the SM allows for the seesaw 
mechanism  \cite{seesaw}
to operate, giving rise to three light and three heavy neutrinos
of Majorana character, as well as leptonic mixing and the possibility
of CP violation in the couplings of the neutrinos to the 
charged leptons. Low energy physics (the decoupling limit)
in this framework, is described by an effective
lefthanded Majorana mass matrix as is the case in models
where only the scalar sector of the SM is enlarged. Yet, the seesaw 
mechanism plays an important r\^ ole in explaining in a 
natural way the ``extreme'' smallness of neutrino masses
when compared to the masses of the other fermions.
Furthermore, CP violation in the decay of the heavy neutrinos 
may lead to a lepton asymmetry which is subsequently 
transformed into a baryon asymmetry, thus providing an
explanation for the observed baryon asymmetry of the 
Universe through leptogenesis. The lepton number asymmetry 
thus produced can be fully parametrised in terms of 
neutrino mass matrices. In flavour models where the number of free
parameters is reduced through the introduction 
of family symmetries or the imposition
of special ans\" atze, it is often possible   
to establish a direct connection between low energy and high energy physics 
in the leptonic sector. 

With the introduction of righthanded neutrino fields, the most
general leptonic mass term after SSB is of the form:
\begin{eqnarray}
{\cal L}_m  &=& -[\frac{1}{2} \nu_{L}^{0T} C m_L \nu_{L}^0+
\overline{{\nu}_{L}^0} m_D \nu_{R}^0 +
\frac{1}{2} \nu_{R}^{0T} C M_R \nu_{R}^0+
\overline{l_L^0} m_l l_R^0] + h. c. = \nonumber \\
&=& - [\frac{1}{2}  n_{L}^{T} C {\cal M}^* n_L +
\overline{l_L^0} m_l l_R^0 ] + h. c.
\label{lrd}
\end{eqnarray}
with the $6 \times 6$ matrix $\cal M $  given by:
\begin{eqnarray}
{\cal M}= \left(\begin{array}{cc}  
{m^*}_L  & m_D \\
m^T_D & M_R \end{array}\right) \label{calm}
\end{eqnarray}
As already explained the appearance of the term 
$\nu_{L}^{0T} C m_L \nu_{L}^0$ would require further 
enlargement of the scalar sector of the Lagrangian. In
what follows, we discuss the minimal seesaw framework where
this term is not present. The terms in $m_D$ are generated 
through Yukawa couplings and therefore cannot be of a scale larger
than the electroweak scale. However,
the terms in $M_R$ are $SU(2) \times U(1)$ invariant, not protected
by any symmetry. Therefore it is natural to assume that their scale is 
much larger than the electroweak scale. The origin
of the term ``seesaw'' is based on the implications of
choosing the scale of $M_R$ much larger than the scale of $m_D$,
as illustrated below. In fact, the existence of these two very
different scales gives rise to two sets of neutrinos of different 
mass scales, one large, of order  of $M_R$, and another one
much suppressed by comparison to the electroweak scale.
This provides a natural explanation for the observed smallness
of neutrino masses.  

After spontaneous symmetry breaking, but before diagonalization
of the fermion mass terms, the leptonic charge gauge interations are 
still diagonal and therefore can be written as:
\begin{equation}
{\cal L}_W = - \frac{g}{\sqrt{2}}  W^+_{\mu} \ 
 \overline{l^0_L} \ \gamma^{\mu} \ \nu^0_{L} +h.c.
\label{16}
\end{equation}
this basis is usually called a weak basis (WB). WB transformations 
are defined as transformations of the fermion fields that leave the
gauge currents flavour diagonal. In the present extension of the SM 
the most general
such transformations are of the form:
\begin{equation}
 l^0_L \longrightarrow U^\prime  l^0_L, \qquad
 \nu^0_L \longrightarrow U^\prime  \nu^0_L, \qquad
 l^0_R \longrightarrow V^\prime  l^0_R, \qquad
 \nu^0_R  \longrightarrow  W^\prime \nu^0_R
\label{WB}
\end{equation}
where $U^\prime$, $V^\prime$ $W^\prime$ are arbitrary unitary matrices.
The lefthanded fields  $l^0_L$, $\nu^0_L$ must transform in the same way
in order to leave the charged weak current of Eq.~(\ref{16}), diagonal.
Since there are no righthanded gauge currents mediated by W, 
in this extension of the SM this constraint does not exist for 
the righthanded fields. 
Physics does not depend on the choice of WB, in particular all WB lead 
to the same fermion masses and mixing. 
Clearly it is always possible to choose 
without loss of generalty a WB where $m_l$ is real
diagonal and positive. In this basis the matrix $V$ that diagonalizes 
$\cal M $ has physical meaning. The diagonalization of the 
matrix $\cal M$ is then performed via the unitary transformation
\begin{equation}
V^T {\cal M}^* V = \cal D \label{dgm}
\end{equation}
where ${\cal D} ={\rm diag} (m_1, m_2, m_3,
M_1, M_2, M_3)$,
with $m_i$ and $M_i$ denoting the physical
masses of the light and heavy Majorana neutrinos, respectively. It is
convenient to write $V$ and $\cal D$ in the following block form:
\begin{eqnarray}
V&=&\left(\begin{array}{cc}
K & G \\
S & T \end{array}\right) ; \label{matv}\\
{\cal D}&=&\left(\begin{array}{cc}
d & 0 \\
0 & D \end{array}\right) . \label{matd}
\end{eqnarray}
The neutrino weak-eigenstates are
related
to the mass eigenstates by:
\begin{eqnarray}
{\nu^0_i}_L= V_{i \alpha} {\nu_{\alpha}}_L=(K, G)
\left(\begin{array}{c}
{\nu_i}_L  \\
{N_i}_L \end{array} \right) \quad \left(\begin{array}{c} i=1,2,3 \\
\alpha=1,2,...6 \end{array} \right)
\label{15}
\end{eqnarray}
and thus the leptonic charged current interactions are given by:
\begin{equation}
{\cal L}_W = - \frac{g}{\sqrt{2}} \left( \overline{l_{iL}} 
\gamma_{\mu} K_{ij} {\nu_j}_L +
\overline{l_{iL}} \gamma_{\mu} G_{ij} {N_j}_L \right) W^{\mu}+h.c.
\label{phys}
\end{equation}
with $K$ and $G$ being the charged current couplings of charged 
leptons to the light neutrinos $\nu_j$ and to the heavy neutrinos 
$N_j$, respectively. From Eqs.~(\ref{dgm}), (\ref{matv}), (\ref{matd})
and for 
\begin{eqnarray}
{\cal M}= \left(\begin{array}{cc}  
 0  & m_D \\
m^T_D & M_R \end{array}\right) \label{zero}
\end{eqnarray}
one obtains:
\begin{eqnarray}
S^\dagger m^T_D K^* + K^\dagger m_D S^*
+ S^\dagger M_R S^* &=&d \label{12a} \\
S^\dagger m^T_D G^* + K^\dagger m_D T^*
+ S^\dagger M_R T^* &=&0 \label{12b} \\
T^\dagger m^T_D G^* + G^\dagger m_D T^*
+ T^\dagger M_R T^* &=&D  \label{12c}
\end{eqnarray}
In the context of seesaw with $M_R$ of a scale $M$, much larger than the
weak scale, $v$, the following relations can be derived from these equations,
valid to an excellent approximation:
\begin{eqnarray}
S^\dagger=-K^\dagger m_D M^{-1}_R \label{13} \\
-K^\dagger m_D \frac{1}{M_R} m^T_D K^* =d \label{14}
\end{eqnarray}
It is clear from 
Eq. (\ref{13}) that S is of order $m_D / M_R$ and therefore is 
very suppressed. Eq. (\ref{14}) is the usual seesaw formula
with the matrix $K$ frequently treated as being equivalent to $U_{PMNS}$,
the Pontecorvo, Maki, Nakagawa, Sakata (PMNS) matrix \cite{pmns}.
Although the block $K$ in Eq. (\ref{matv}) is not a unitary matrix 
its deviations from unitarity are of the order $m^2_D / M^2_R$.
It is from  Eq. (\ref{14}) that the low energy physics of
the leptonic sector is derived. The decoupling limit
corresponds to an effective theory with only lefthanded neutrinos
and a Majorana mass matrix, $m_{eff}$ defined as:
\begin{equation}
m_{eff} = - m_D \frac{1}{M_R} m^T_D \label{meff}
\end{equation} 
showing that for $m_D$ of the order of the electroweak scale and $M_R$
of the scale of grand unification, the smallness of light
neutrino masses is a natural consequence of the seesaw 
mechanism \cite{seesaw}.
From the relation ${\cal M}^* V = V^*  \cal D $ and taking into
account the zero entry in ${\cal M}$ one derives the following
exact relation
\begin{equation}
G=m_D T^* D^{-1} \label{exa}
\end{equation} 
This equation plays an important r\^ ole in the connection between
low energy and high energy physics in the leptonic sector,
and shows explicitly that the suppression in the matrix $G$ 
is of the same order of the suppression in $S$ as required
by unitarity of the matrix $V$.\\

There are in the literature excellent reviews \cite{Mohapatra:1998rq}
\cite{Cheng:1985bj} on the seesaw mechanism, showing 
explicitly that under this mechanism the resulting 
physical fermions are  in general Majorana spinors. 
The left and the righthanded components of Majorana spinors 
are not independent. Out of two independent 
spinor components $\Psi_L$ and  $\Psi_R$ one can form a Dirac spinor:
\begin{equation}
\Psi = \Psi_L + \Psi_R
\label{dir}
\end{equation}
or two Majorana spinors:
\begin{eqnarray}
\chi =  \Psi_L +  \Psi^c_L  \qquad  \Psi^c_L \equiv C \  
{\overline   \Psi_L}^T \\
\omega = \Psi_R +  \Psi^c_R  \qquad  \Psi^c_R \equiv C \   
{\overline   \Psi_R}^T 
\end{eqnarray}
Majorana spinors have the property of being self-conjugate, that is:
\begin{equation}
\chi^c = \chi, \qquad \omega^c = \omega \qquad (\Psi^c \equiv   
C \  {\overline   \Psi}^T = C \gamma^T_0 \  \Psi^\ast )
\end{equation}
The most general definition of a Majorana spinor allows for a relative
phase in the components of $\chi$ and of $\omega$ which would
manifest itself  in the self conjugate condition.

\section{CP violation in the quark and lepton sectors}

\subsection{The quark sector}
A thorough discussion of CP violation in the 
SM and  in some of its extensions can be found in 
\cite{Branco:1999fs}. Here we only address a selected number of
topics.

Gauge invariance does not constrain the flavour structure of 
Yukawa interactions. As a result, in the SM quark masses and mixing
are arbitrary.
It has been shown that gauge theories with fermions, but without
scalar fields, do not break CP symmetry \cite{Grimus:1995zi}.
A Higgs doublet is used in the SM to break both the gauge symmetry 
and generate fermion masses through Yukawa  interactions.
Yukawa couplings have the special feature of being the only
couplings of the SM which can be complex. All other couplings are
constrained to be real, by hermiticity. This is the essential
reason why, in the context of the SM, Yukawa couplings play a crucial
r\^ ole in generating CP violation. Indeed CP violation in
the SM can only arise from the simultaneous presence of Yukawa 
and gauge interactions. For three or more fermion generations
CP violation can be broken at the Lagrangian level.
In the SM where a single 
Higgs doublet is introduced, it is not possible to have spontaneous
CP violation since any phase in the vacuum expectation value
(vev) of the neutral Higgs can 
be eliminated by rephasing the Higgs field. Furthermore,
in the SM it is also not possible to violate CP explicitly 
in the Higgs sector
since gauge invariance together with renormalizability restrict
the potential:
\begin{equation}
V = - \mu^2 \phi^\dagger \phi + \lambda \left( 
\phi^\dagger \phi \right)^2 + \mbox{h.c.}
\label{vhig}
\end{equation}
to have only quadratic and quartic couplings and
hermiticity constrains both of these terms to be real. 

The Yukawa interactions for the quark sector can be written as:
\begin{equation}
{\cal L}_Y \mbox{(quarks)}
= g_{ij}\ \overline{{q_L}^0_i}\  \widetilde{\phi}\  {u^0_R}_j +
f_{ij} \ \overline{{q_L}^0_i}\  \phi\  {d^0_R}_j + \mbox{h.c.}
\label{lyu}
\end{equation}
with $q^0_L$ the lefthanded quark doublets and $\widetilde{\phi} = 
i \tau_2 {\phi}^*$. After SSB the following quark mass terms
are generated
\begin{equation}
{\cal L}_m \mbox{(quarks)} = - \overline{u^0_L} \ m_u \ u^0_R -
 \overline{d^0_L} \ m_d \ d^0_R  +  \mbox{h.c.}
\label{qmas}
\end{equation}
We are still in a WB, so the charged current is diagonal, of the form:
\begin{equation}
{\cal L}_W \mbox{(quarks)}= - \frac{g}{\sqrt{2}}\   W^+_{\mu} \ 
 \overline{u^0_L} \ \gamma^{\mu} \ d^0_{L} +\mbox{h.c.}
\label{Wqua}
\end{equation}
The mass matrices are general complex matrices and may be diagonalized
through a bi-unitary transformation:
\begin{equation} 
u_L = U^u_L \ u^0_L, \qquad u_R = U^u_R \ u^0_R, \qquad
d_L = U^d_L \ u^0_L, \qquad d_R = U^d_R \ d^0_R
\label{biu}
\end{equation}
such that:
\begin{eqnarray} 
{U^u_L}^\dagger \ m_u \  U^u_R = \mbox{diag}\  (m_u, \ m_c, \ m_t), 
\label{abc} \\
{U^d_L}^\dagger \ m_d \  U^d_R = \mbox{diag}\  (m_d, \ m_s, \ m_b).
\label{efg}
\end{eqnarray}
After this transformation the charged currents are no longer diagonal.
In terms of quark mass eigenstates the charged currents are now given by:
\begin{equation}
{\cal L}_W \mbox{(quarks)} = - \frac{g}{\sqrt{2}}\   W^+_{\mu} \ 
 \overline{u_L} \ \gamma^{\mu} V_{CKM}\ d_{L} +\mbox{h.c.}
\label{ckm}
\end{equation}
where $V_{CKM} = {U^u_L}^\dagger \ U^d_L $, denotes the 
Cabibbo--Kobayashi--Maskawa (CKM) matrix. The appearance of a nontrivial CKM 
matrix in the charged currents reflects the fact that the Hermitian 
matrices $H_u$ and $H_d$ defined as:
\begin{equation}
H_u = m_u \ {m_u}^\dagger, \qquad H_d = m_d \ {m_d}^\dagger
\label{hem}
\end{equation}  
 are in general diagonalized by different unitary matrices:
\begin{eqnarray}
{U^u_L}^\dagger \ H_u \  U^u_L = \mbox{diag}\  (m^2_u, \ m^2_c, \ m^2_t), \\
{U^d_L}^\dagger \ H_d \  U^d_L = \mbox{diag}\  (m^2_d, \ m^2_s, \ m^2_b).
\end{eqnarray}  
In fact, of the four unitary matrices appearing in Eqs.~(\ref{abc})
and (\ref{efg}) only the matrices $U^u_L$ and  $U^d_L$ play a r\^ ole
in generating $V_{CKM}$ which encodes the physical quark mixing and
CP violation. In the SM, one can use the freedom to make WB
transformations to choose, without loss of generality, a basis where 
$m_u$, $m_d$ are hermitian. Furthermore, one may also choose without
loss of generality a basis where $m_u$ (or $m_d$) is diagonal
and  $m_d$ (or $m_u$) are hermitian.

Given a Lagrangean, obtained for instance from model building,
one may ask whether or not it violates CP.
In the context of the SM one may always investigate the CP
properties by going to the physical basis and analysing the CKM matrix.
However, it may be useful to try to answer the same question 
still in a WB without requiring cumbersome changes of basis. 
In this case
the relevant information is contained in the matrices $m_u$ and
$m_d$. The general method allows for construction of weak basis invariants 
which have to vanish in order for CP symmetry to hold 
and was first proposed in \cite{Bernabeu:1986fc} for the Standard Model.
Weak basis invariant conditions relevant for CP violation in the
leptonic sector were later developed and
are discussed in some detail in the next subsection. 
This approach has been widely applied in the literature 
to study CP violation in many other scenarios  \cite{many1}.
The strategy is to apply the most general CP transformation
for fermion fields in a WB, i.e., leaving the gauge interaction invariant:
\begin{eqnarray}
{\rm CP} u^0_L ({\rm CP})^{\dagger}&=&U^\prime
\gamma^0  C~ \overline{u^0_L}^T;  \quad
{\rm CP} u^0_R({\rm CP})^{\dagger}=V^\prime \gamma^0  C~ \overline{u^0_R}^T
\nonumber \\
{\rm CP} d^0_L ({\rm CP})^{\dagger}&=&U^\prime \gamma^0 C~
\overline{d^0_L}^T; \quad
{\rm CP} d^0_R ({\rm CP})^{\dagger}=W^\prime \gamma^0  C~
\overline{d^0_R}^T \label{cp}  \\
{\rm CP} W^+_{\mu} ({\rm CP})^{\dagger} &=& - (-1)^{\delta_{0\mu}}  
W^-_{\mu} \nonumber
\end{eqnarray}
where $U^\prime$, $V^\prime$, $W^\prime$ are unitary matrices 
acting in flavour space not related to those introduced in the
previous section. This transformation can be viewed as a
combination of the CP
transformation of a single fermion field with a WB 
transformation \cite{Ecker:1981wv}.
Invariance of the mass terms under the above CP transformation,
requires that the following relations have to be 
satisfied \cite{Bernabeu:1986fc}:
\begin{eqnarray}
U^{\prime \dagger} m_u V^\prime &=& {m_u}^* \label{cpmu} \\
U^{\prime \dagger} m_d W^\prime &=& {m_d}^* \label{cpmd} 
\end{eqnarray}
It can be easily seen that if there are unitary matrices 
$U^\prime$, $V^\prime$, $W^\prime$ satisfying 
Eqs. (\ref{cpmu}), (\ref{cpmd}) in one particular WB, then 
a solution exists for any other WB. It is also clear that for 
$m_u$ and $m_d$ real these conditions are trivially satisfied
for $U^\prime$, $V^\prime$, $W^\prime$ equal to the identity
matrix. This shows that the existence of CP violation in the SM 
does require Yukawa couplings to be complex. In this form these 
conditions are not yet very useful since at this stage one just 
replaced the requirement of diagonalizing the mass matrices by 
that of finding these three unitary matrices. However, combining 
these equations in such a way as to end up with
similarity transformations one may be rid of the 
unitary matrices and derive necessary and sufficient conditions
for CP invariance, expressed in terms of invariants (traces, determinants).
In this way one may derive
the following condition
\begin{equation}
{\rm tr} \left[ H_u,   H_d \right]^3 = 0
\label{lll}
\end{equation}
which is a necessary and sufficient condition for CP invariance in the 
SM with three generations \cite{Bernabeu:1986fc}. 
This invariant condition can be applied
in any WB, as was already stressed. It can also be expressed in terms
of physical quantities in the form:
\begin{eqnarray}
{\rm tr} \left[ H_u,   H_d \right]^3 = 6i\ (m^2_t -  m^2_c) \
(m^2_t -  m^2_u) \ (m^2_c -  m^2_u) \times \nonumber \\
\ ( m^2_b -  m^2_s) \  ( m^2_b -  m^2_d) \ 
(m^2_s -  m^2_d) \ {\rm Im} \left( 
V_{12}V_{23}V^*_{13}V^*_{22} \right)
\label{uuu}
\end{eqnarray}
where  $V_{ij}$ denote the entries of $V_{CKM}$.
For three generations the condition of Eq.~(\ref{lll}) is equivalent to:
\begin{equation}
{\rm det} \left[ H_u,   H_d \right] = 0
\label{trr}
\end{equation} 
This expression was first given in  \cite{Jarlskog:1985ht}.
Note that the condition of Eq.~(\ref{trr}) only applies to
an odd number of generations, while   Eq.~(\ref{lll}) 
is a necessary condition for CP invariance, in any number of
generations. \\

In the physical basis, i.e. after diagonalization of the
quark mass matrices,
the only terms of the SM Lagrangian that 
may violate CP are the charged current interactions
which are parametrised by the $V_{CKM}$ matrix. 
CP can only be violated if $V_{CKM}$ is a complex matrix.
However, not all of its phases have physical meaning since there 
is freedom to rephase the quark fields:
\begin{equation}
u_i = e^{i\alpha_i} u^\prime_i, \qquad d_j = e^{i\beta_j} d^\prime_j
\qquad V^\prime_{ij} = e^{i(\beta_j - \alpha_i)} V_{ij}
\label{mmm}
\end{equation}
This allows to eliminate 
five phases out of the nine that may in principle be present 
in $V_{CKM}$. Physically meaningful quantities must be invariant under
rephasing of the fields. Rephasing invariants involve products of
several elements of $V_{CKM}$ and its complex conjugate. 
In the absence of zero entries all rephasing invariants may
be expressed in terms of the simplest ones which are the 
moduli of matrix elements $|V_{ij}|$ and the terms called
quartets which are of the form:
\begin{equation}
Q_{ijkl} \equiv V_{ij}V_{kl}V^*_{il}V^*_{kj}
\label{qua}
\end{equation}
for $i\neq k$ and $j\neq l$.

In the physical  
basis the most general CP transformation for the quarks and 
for the $W$ boson are of the form  \cite{Branco:1999fs}:
\begin{eqnarray}
{\rm CP} \ u_i \ ({\rm CP})^{\dagger}&=& e^{(i{\sigma}_i)} \ 
\gamma^0  C~ \overline{u_i}^T;  \quad
{\rm CP} \  d_j \ ({\rm CP})^{\dagger} =  e^{(i{\kappa}_j)} \ 
\gamma^0 C~ \overline{d_j}^T; \\
{\rm CP} \ W^+_{\mu} \ ({\rm CP})^{\dagger} &=& -   e^{(i{\rho}_w)}
W^-_{\mu} \nonumber
\end{eqnarray}
Note that the CP transformation no longer mixes fermion generations  
since the quark mass terms are already diagonal
and there is no mass degeneracy. 
Invariance of the Lagrangian under this transformation requires
\begin{equation}
 V^*_{ij} = e^{i({\rho}_w + \kappa_j - \sigma_i)} V_{ij}
\end{equation}
This equation can always be made to hold if one considers a single
matrix element of $V_{CKM}$, because the CP transformation phases
$\sigma_i$, ${\kappa}_j$ and ${\rho}_w$,  are arbitrary. Obviously for
a real $V_{CKM}$ this condition is trivially verified. However, 
imposing this condition on each element of $V_{CKM}$   simultaneously
forces the quartets and all other rephasing-invariant functions of
$V_{CKM}$, to be real. In general, \cite{Branco:1999fs} {\it there is
CP violation in the SM if and only if any of the rephasing-invariant
functions of the CKM matrix is not real}.

It can be easily shown that as a consequence of the orthogonality
of any pair of different rows or columns of the CKM matrix the 
imaginary parts of all quartets are equal up to their sign.
Let us count the number of independent parameters in $V_{CKM}$.
An $n \times n$ unitary matrix has $n^2$ independent parameters.
Taking into account that 
$(2n-1)$ phases can be removed from $V_{CKM}$, through rephasing of the
$2n$ quark fields (note that an overall rephasing of quark fields does
not affect $V_{CKM}$) the number of physical parameters in  $V_{CKM}$
is:
\begin{equation}
N_{\mbox{param}} = n^2 - (2n-1) = (n-1)^2
\end{equation}
An  orthogonal  $n \times n$ matrix $O(n)$ is parametrised by 
$n(n-1)/2$ rotation angles which
are sometimes called Euler angles. An unitary matrix is a
complex extension of an orthogonal matrix. Therefore out of the 
$N_{\mbox{param}}$ parameters of  $V_{CKM}$,
\begin{equation}
N_{\mbox{angle}} = \frac{1}{2} n (n-1)
\end{equation}
should be identified with rotation angles. The remaining
\begin{equation}
N_{\mbox{phase}} = N_{\mbox{param}} - N_{\mbox{angle}} =
 \frac{1}{2}(n-1)(n-2)
\end{equation}
parameters of $V_{CKM}$ are physical phases.
For $n=3$ there is one phase and three mixing angles, and
$V_{CKM}$ can be written as: 
\begin{small}
\begin{eqnarray}
V_{CKM} =\left(
\begin{array}{ccc}
c_{12} c_{13} & s_{12} c_{13} & s_{13} e^{-i \delta}  \\
-s_{12} c_{23} - c_{12} s_{23} s_{13}   e^{i \delta}
& \quad c_{12} c_{23}  - s_{12} s_{23}  s_{13} e^{i \delta} \quad 
& s_{23} c_{13}  \\
s_{12} s_{23} - c_{12} c_{23} s_{13} e^{i \delta}
& -c_{12} s_{23} - s_{12} c_{23} s_{13} e^{i \delta}
& c_{23} c_{13} 
\end{array}\right)
\label{std}
\end{eqnarray}
\end{small}
where $c_{ij} \equiv \cos \theta_{ij}\ , \ s_{ij} \equiv \sin \theta_{ij}\ $
and $\delta$ is the only phase. $\delta$  is called a Dirac-type phase
because it is the Dirac character of the quarks that allows to
rephase away all other phases leaving only $\delta$. 
This is the so-called standard parametrisation of $V_{CKM}$ \cite{Yao:2006px}.
The mechanism just described for CP violation is the 
Kobayashi-Maskawa (KM) mechanism. In the SM this is the 
only source of CP violation. Notice that 
$\delta$ is not a rephasing invariant quantity, it is only
meaningful within a given parametrisation.

A particularly useful  phase convention for $V_{CKM}$ only in terms of
rephasing invariant quantities is  \cite{Branco:1999fs}:
\begin{equation}
V_\mathrm{CKM}=\left(
\begin{array}{ccc}
\left| V{ud}\right| &\left| V{us} \right| e^{i\chi^{\prime }} 
& \left| V{ub} \right| e^{-i\gamma } \\
-\left| V{cd}\right| &\left|  V{cs}\right| &\left| V{cb}\right| \\
\left| V{td}\right| e^{-i\beta } & -\left| V{ts}\right| 
e^{i\chi } &\left|  V{tb}\right|
\end{array}
\right)  \label{VCKM1}
\end{equation}%
where the CP-violating phases introduced in Eq. (\ref{VCKM1})
are defined by:
\begin{equation}
\begin{array}{ccc}
\beta =\arg \left( -V_{cd} V^{\ast}_{cb} V^{\ast}_{td}V_{tb}\right) & , &
\gamma =\arg \left( -V_{ud}V^{\ast}_{ub}V^{\ast}_{cd} V_{cb}\right)~, \\
\chi =\arg \left( -V_{ts}V^{\ast}_{tb}V^{\ast}_{cs}V_{cb}\right) & , &
\chi ^{\prime }=\arg \left( -V_{cd}V^{\ast}_{cs}V^{\ast}_{ud} V_{us}\right)~.
\end{array}
\label{CPphas0}
\end{equation}
Without imposing the
constraints of unitarity, the four rephasing invariant phases,
together with the nine moduli are all the independent physical
quantities contained in $V_\mathrm{CKM}$. In the SM, where unitarity
holds, these quantities are related by a series of exact relations
which provide a stringent test of the SM  \cite{Botella:2002fr}.

\begin{figure}[htb]
\begin{center}
\mbox{\epsfig{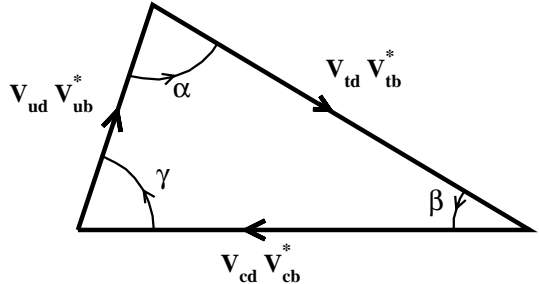}}
\end{center}
\caption{The Standard Model unitary triangle.
\label{triangle1}}
\end{figure}

Unitarity of $V_{CKM}$ implies orthogonality of rows and columns.
Let us consider the orthogonality between  the first and third
column:
\begin{equation}
V_{ud}V^{\ast}_{ub} +V_{cd}V^{\ast}_{cb} +  V_{td}V^{\ast}_{tb} = 0
\label{eq60}
\end{equation}
This equation may be interpreted as representing a triangle in
the complex plane. One may in principle build in this way 
three different triangles from orthogonality of columns and three 
other triangles from orthogonality of rows. Out of the six unitarity 
triangles, only two have sides of comparable size, the one 
corresponding to Eq.~(\ref{eq60}) and the one corresponding to
orthogonality of the first and the third rows. A remarkable feature 
of the unitarity triangles is the fact that all of them have the same 
area. Phenomenologically 
the most interesting triangle is the one depicted 
in Fig.~\ref{triangle1}, corresponding to  Eq.~(\ref{eq60})
which is often refered to
in the literature as {\it the} unitary triangle. 
The angle $ \alpha$, represented in the figure, is defined as 
$\alpha \equiv =\arg \left( -V_{td} V^{\ast}_{tb} V^{\ast}_{ud}V_{ub}\right)$
and obeys by definition the relation
$ \alpha =\pi -\beta -\gamma $. Rephasing of the CKM matrix,
as in Eq.~(\ref{mmm}), rotates 
the triangle as a whole, since under rephasing:
\begin{equation}
 V^\prime_{ij} V^{\prime \ast}_{ik} = 
e^{i(\beta_j - \beta_k)} V_{ij} V^{\ast}_{ik}
\end{equation}
therefore the orientation of the triangle is physically
meaningless. However, the shape of the triangle remains unchanged
because both its inner angles and the lengths of its sides are
rephasing invariant. \\

In the SM neutrinos are massless 
and there is no leptonic mixing. Furthermore in the SM there are no 
flavour changing neutral currents (FCNC) at tree level
neither mediated by the Z nor by neutral scalar fields. Therefore
the only source of CP violation in the SM is the KM mechanism just
described.

\subsection{The Lepton sector}
In the previous section the seesaw mechanism was explained,
working on the WB where the mass matrix $m_l$ was chosen to be 
real and diagonal. In order to discuss CP violation in this framework
it is useful to adopt as a starting point
a WB were, in addition to $m_l$, the matrix $M_R$
is also chosen to be real and diagonal. It should be clear from 
Eqs.~(\ref{lrd}) and (\ref{WB}) that this is indeed a possible 
choice of WB. In this case, since we are working 
in a framework where $m_L$ is not introduced,
all phases appear in the matrix $m_D$ and the leptonic mass
matrix becomes:
\begin{eqnarray}
{\cal L}_m  &=& 
- \overline{{\nu}_{L}^0} m_D \nu_{R}^0 -
\frac{1}{2} \nu_{R}^{0T} C D_R \nu_{R}^0 -
\overline{l_L^0} d_l l_R^0 + h. c. 
\label{xyz}
\end{eqnarray}
The matrix $D_R$ coincides with the matrix $D$ of Eq.~(\ref{matd})
up to negligible corrections. This can be seen from Eq.~(\ref{12c})
since the matrix $G$ is very suppressed and $m_D$ is of order
much smaller than $M_R$. In this WB, the exact relation given by
Eq.~(\ref{exa}) is very well approximated by:
\begin{equation}
G=m_D D^{-1} \label{nexa}
\end{equation} 
The matrix $d_l$ is also diagonal and contains
the masses of the charged leptons, therefore 
we could have considered dropping the $0$ upper index for
these fields since these are already physical fields up to 
phase redefinitions. The matrix $m_D$ is perfectly general, it 
contains nine real parameters -- the moduli of each entry -- and nine 
phases. However, there is still freedom to rotate away three of these
phases through the rephasing of the ${{\nu}_{L}^0}$ fields. 
These phases would appear in ${\cal L}_W$ of Eq.~(\ref{16}), 
however, they can be eliminated by rotating $l^0_L$. Finally 
a rotation of the fields $l^0_R$ would also eliminate these
phases from $d_l$. We are 
thus left with six real parameters in $D_R$ and $d_l$ plus nine
real parameters in $m_D$ and six phases. The following special possible 
parametrisations for $m_D$:
\begin{equation}
m_D = U Y_\triangle \qquad \mbox{or} 
\qquad m_D = U H
\label{usef}
\end{equation}
where $U$ is an unitary matrix, $H$ is an Hermitian matrix and
$Y_\triangle$ is a lower triangular
matrix, have revealed themselves particularly useful in model building.
The number of parameters
in this WB equals the number of physical parameters -- in the form 
of masses and mixing --
obtained after diagonalization of the mass matrices.
In this case there are the nine masses of the 
three charged leptons, the three light neutrinos and the three heavy 
neutrinos, together with six mixing angles required to parametrise
the $3 \times 6$ physical block ($K, G$) of the $6 \times 6$ 
unitary matrix $V$ \cite{Branco:1986my} as well as  six phases 
\cite{Endoh:2000hc}. In general, with $m_L$ different from zero 
one would have twelve independent phases \cite{Branco:1986my} in the 
mixing matrix. Is is easy to understand why having $m_L$
equal to zero reduces the number of CP violating phases. Notice that 
$m_L$ is in general a complex symmetric matrix and therefore would
have six phases in the case of three generations. Once $m_L$ is equal to zero,
from $ {\cal M}^*  = V^* {\cal D} V^\dagger $,
the zero entry in the upper left block of $\cal M$ implies:
\begin{equation}
K^* d K^\dagger + G^* D G^\dagger =0. \label{krd}
\end{equation}
providing additional constraints for the matrices $K$ and $G$
beyond those derived from unitarity of the matrix $V$.

It is quite straightforward to determine the number of independent
CP restrictions, by making use of the WB basis chosen above,
for the general case of $n$ generations
\cite{Branco:2001pq}. Invariance of the mass terms under the
most general CP transformation which leaves the gauge interaction 
invariant:
\begin{eqnarray}
{\rm CP} l^0_L ({\rm CP})^{\dagger}&=&U^\prime
\gamma^0  C~ \overline{l^0_L}^T;  \quad
{\rm CP} l^0_R({\rm CP})^{\dagger}=V^\prime \gamma^0  C~ \overline{l^0_R}^T
\nonumber \\
{\rm CP} \nu^0_L ({\rm CP})^{\dagger}&=&U^\prime \gamma^0 C~
\overline{\nu^0_L}^T; \quad
{\rm CP} \nu^0_R ({\rm CP})^{\dagger}=W^\prime \gamma^0  C~
\overline{\nu^0_R}^T \label{cpcp}  
\end{eqnarray}
requires that the following relations have to be satisfied:
\begin{eqnarray}
{W^\prime}^T D_R W^\prime &=&-D_R^* \label{cpM} \\
{U^\prime}^{\dagger} m_D W^\prime&=& {m_D}^* \label{cpm} \\
{U^\prime}^{\dagger} d_l V^\prime&=& {d_l}^* \label{cpml}
\end{eqnarray}
From Eq.~(\ref{cpM})  ${W^\prime}$ is  constrained to be of the form
\begin{equation}
W^\prime={\rm diag} \left(\exp(i\alpha_1), \exp(i\alpha_2),... 
\exp(i\alpha_n)
\right) \label{expw}
\end{equation}
and the $\alpha_i$ have to satisfy:
\begin{equation}
\alpha_i=(2 p_i +1) \frac{\pi}{2} \label{ais}
\end{equation}
with $p_i$ integer numbers. Multiplying Eq.~(\ref{cpml}) by its Hermitian
conjugate, one concludes that $U^\prime$ has to be of the form:
\begin{equation}
U^\prime ={\rm diag} \left(\exp(i\beta_1), \exp(i\beta_2),... \exp(i\beta_n)
\right) \label{expu}
\end{equation}
where $\beta_i$ are arbitrary phases. From Eqs.~(\ref{cpm}),
(\ref{expw}), (\ref{expu}) it follows then
that CP invariance constrains the matrix $m_D$ to satisfy :
\begin{equation}
{\rm arg}(m_D)_{ij}=\frac{1}{2}(\beta_i-\alpha_j) \label{arg}
\end{equation}
Note that the $\alpha_i$ are fixed by Eq.~(\ref{ais}), up to discrete
ambiguities. Therefore CP invariance constrains the matrix $m_D$
to have only $n$ free phases $\beta_i$. Since $m_D$ is in an arbitrary
matrix, with
$n^2$ independent phases, it is clear that there are $n^2-n$ independent
CP restrictions. This number equals, of course, the number of independent
CP violating phases which appear in general in this model. In the WB
which
we are considering, these phases appear as $n(n-1)$ phases which cannot
be removed from $m_D$. 
It should be pointed out that it is also possible
to generate neutrino masses in such a framework without requiring
the number of righthanded and lefthanded neutrinos to be equal.
When the number of righthanded neutrinos is
$n^\prime$ different from $n$, the matrix $m_D$ has dimension
$n n^\prime$, in this case the number of CP violating phases is equal
to $n n^\prime-n$.

In the context of seesaw, CP violation occurs both at 
low and high energies. It is clear from Eq.~(\ref{phys}) that
CP violation at high energies will manifest itself in the decays
of heavy neutrinos. These decays provide a possible source for 
the generation in the early Universe of the baryon asymmetry of the 
Universe (BAU) through leptogenesis \cite{Fukugita:1986hr}. 
A detailed analysis on the present theoretical and experimental
situation  in neutrino physics and on where it is going in the 
future is done in Ref.~ \cite{Mohapatra:2005wg}

\subsubsection{CP violation at low energies}

We start by summarising what is presently known about neutrino
masses and leptonic mixing. 
For a detailed account of the present experimental status of
neutrino physics see the contribution of David L.Wark
``Experimental neutrino physics'' in this volume. 
It is by now experimentally established 
that neutrinos have masses and that there is mixing in the
leptonic sector.  At low energies, only the first term of 
Eq.~(\ref{phys}) involving charged leptons and light neutrino
couplings to the W boson is relevant, since heavy neutrinos in 
the seesaw framework are expected to have masses that may
be of order $10^{13}$ Gev or even larger. Such heavy
neutrinos cannot be produced at present colliders and
would have decayed in the early Universe.

In the seesaw framework,  
described before, $m_{eff}$ given by Eq.~(\ref{meff})
is an effective Majorana mass matrix and the mixing matrix $K$
can be treated as the unitary matrix that diagonalises $m_{eff}$
in Eq.~(\ref{14}). Deviations from unitarity cannot be 
experimentally observed at the level predicted in this framework. 
With the usual conventions, where the Majorana mass term is given by 
$ \nu_{L}^{0T} C m_\nu \nu_{L}^0$, and the PMNS matrix defined by
$U^T_{PMNS} m_{\nu} U_{PMNS} =$ diag$(m_1, m_2, m_3)$,
we have the following correspondence:
\begin{equation}
m_{eff} =  m^*_\nu \qquad \mbox{and} \qquad U_{PMNS} = K^* \label{corr} 
\end{equation}
This effective low energy physics corresponds to integrating 
out the heavy neutrinos. Majorana mass terms are symmetric by 
construction. In fact, anticommutation of fermion fields
together with the property that $C$ is an antisymmetric matrix,
$C^T = -C$, which follows from its definition, leads to
$ \nu_{Li}^{0T} C  \nu_{Lj}^0 = \nu_{Lj}^{0T} C  \nu_{Li}^0$.

In the physical basis the mass terms and the leptonic charged 
currents in the low energy effective theory are of the form:
\begin{equation}
{\cal L}^{phys}_{eff} = \nu^T_{L} C \ d \ \nu_{L} +
\overline{l_L}\  d_l \  l_R +  \frac{g}{\sqrt{2}}  \overline{l_{iL}} 
\gamma_{\mu} U_{ij} {\nu_j}_L W^{+\mu} + h. c. 
\end{equation}
with $d$ and $d_l$ diagonal real and positive matrices.
For simplicity, we have dropped the index PMNS in the mixing matrix $U$. 
The $3 \times 3$ unitary matrix $U$ is in general parametrised by 
six phases and three mixing angles. Three of these phases can be 
factored out  to the left and rotated away through the redefinition 
of the charged leptons $l_L$. The phases thus appearing in $d_L$ 
can be eliminated by the simultaneous redefinition of the fields 
$l_R$. Another two of the six phases of $U$ can be factored out to
the right. However, in this case, these two phases are physical,
since rotating them away from $U$ simply corresponds to transferring
then to the mass term of the light neutrinos. This is an important 
difference from the quark sector resulting from the fact that
in the seesaw framework neutrinos have Majorana masses and
are Majorana particles,  unlike quarks. These factorizable phases 
that cannot be removed from the theory are called Majorana
phases. As a result, in the seesaw framework, in the effective low 
energy theory, there are additional sources for 
CP violation beyond the Kobayashi-Maskawa mechanism of the 
hadronic sector,  one is a Dirac 
type CP violating phase appearing in the leptonic sector,
analogous to the one of the quark sector, together
with two additional  Majorana type phases. The PMNS
matrix may be parametrised  by a matrix of 
the same form as the one given in Eq.~(\ref{std}) for $V_{CKM}$,
multiplied by a diagonal matrix $P$ with two phases:
\begin{equation}
P=\mathrm{diag} \ (1,e^{i\alpha}, e^{i\beta})
\end{equation}
with $\alpha $ and $\beta$ denoting phases associated to the
Majorana character of neutrinos.

It is important to notice that including  a phase
$\gamma$ in the mass term is equivalent to work with a Majorana
spinor field $\chi_\nu$ defined by:
\begin{equation}
\chi_\nu = \nu_L + e^{i\gamma} \nu^c_L
\end{equation}
since in this case the Majorana mass term is given by:
\begin{eqnarray}
{\cal L}_M (\chi_\nu) \equiv m_M \overline{\chi_\nu} \chi_\nu \ 
&=& \  m_M  \left( \overline{\nu_L} + e^{-i\gamma} \ \overline{\nu^c_L}
\right) \left( \nu_L + e^{i\gamma} \ \nu^c_L \right) \ = 
\nonumber \\
&=& m_M \ e^{-i\gamma} \ \nu^T_L C \nu_L + h. c.
\end{eqnarray}
A Majorana spinor defined as $\chi_\nu$ obeys the following self-conjugate 
relation:
\begin{equation}
\chi^c_\nu =  \nu^c_L +  e^{-i\gamma} \nu_L \ = \  
e^{-i\gamma} \chi_\nu 
\end{equation}

Experimentally it is not yet known whether any of the three CP violating 
phases of the leptonic sector is different from zero.
The current experimental bounds on neutrino masses and leptonic
mixing are \cite{Yao:2006px}:
\begin{eqnarray}
\Delta m^2_{21} & = & 8.0 ^{+0.4}_{-0.3} \times 10^{-5}\  {\rm eV}^2 \\
\sin^2 (2 \theta_{12}) & = & 0.86 ^{+0.03}_{-0.04} \\ 
|\Delta m^2_{32}| & = & (1.9 \ \  \mbox{to} \ \  3.0) \times 10^{-3}\  
{\rm eV}^2 \\
\sin ^2 ( 2 \theta_{23}) & > & 0.92 \\ 
\sin ^2  \theta_{13} &  < & 0.05 
\end{eqnarray} 
with $\Delta m^2_{ij} \equiv m^2_j - m^2_i$. The angle $ \theta_{23} $ 
may be maximal, meaning $45^{\circ}$, whilst  
$ \theta_{12} $ is already known to deviate from this value. At the moment, 
there is only an experimental upper bound on the angle $ \theta_{13}$.
All present data is consistent with the
Harrison, Perkins and Scott (HPS) mixing matrix \cite{Harrison:2002er}:
\begin{equation}
\left[ 
\begin{array}{ccc}
 \frac{2}{\sqrt 6}  & \frac{1}{\sqrt 3} & 0 \\ 
 - \frac{1}{\sqrt 6}  &  \frac{1}{\sqrt 3} &  \frac{1}{\sqrt 2} \\ 
  - \frac{1}{\sqrt 6} & \frac{1}{\sqrt 3} & - \frac{1}{\sqrt 2}
\end{array}
\right]
\label{scott}
\end{equation}
which exhibits a so-called tri-bimaximal mixing. 
There have been various
attempts at introducing family symmetries leading to this structure. 
Some examples can be found in Ref.~\cite{tribi}.

It is also not yet known
whether the ordering of the light neutrino masses is normal, i.e,
$m_1<m_2<m_3$ or inverted $m_3<m_1<m_2$. The scale of the neutrino
masses is also not yet established. Direct kinematical limits from
Mainz  \cite{Kraus:2004zw} and Troitsk \cite{Lobashev:1999tp}
place an upper bound on $m_{\beta}$ defined as:
\begin{equation}
m_{\beta} \equiv \sqrt{\sum_{i} |U_{ei}|^2 m^2_i}
\end{equation}
given by $m_{\beta} \leq 2.3$ eV (Mainz), 
$m_{\beta} \leq 2.2$ eV  (Troitsk). The forthcoming KATRIN experiment
\cite{Osipowicz:2001sq} is expected to be sensitive to
$m_{\beta} > 0.2$ eV. 
The spectrum may vary from extreme 
hierarchy, between the two lightest neutrino masses, to three 
quasidegenerate masses. Examples of the possible extreme cases are: 
\begin{equation}
m_1 \sim 0 \ \  (\mbox{or e.g.} \sim 10^{-6} \mbox{eV}),\ \   
m_2 \simeq 9 \times 
10^{-3} \mbox{eV}, \ \   m_3 \simeq 5 \times 10^{-2} \mbox{eV}
\end{equation}
corresponding to normal spectrum,  hierarchical, or else:
\begin{equation}
m_3 \sim 0 \ \  (\mbox{or e.g.} \sim 10^{-6} \mbox{eV}), \ \ 
m_1 \simeq m_2  \simeq  0.05  \mbox{eV}
\end{equation}
corresponding to inverted spectrum, hierarchical, or else:
\begin{equation}
m_1 \simeq 1 \mbox{eV}, \ \   m_2 \simeq 1 \mbox{eV}, \ \   
m_3 \simeq 1 \mbox{eV}
\end{equation}
corresponding to almost degeneracy.

The limit of exact mass degeneracy of Majorana neutrinos
was studied in \cite{Branco:1998bw} 
where it was shown that it has the remarkable feature of allowing 
for the existence of mixing and CP violation. In the exact degeneracy 
limit the leptonic mixing matrix is parametrized 
by only two angles and one 
phase and there is no Dirac type CP violation in the leptonic sector.
However, there may be Majorana-type CP violation.

It is possible to obtain information on the absolute scale of
neutrino masses from the study of the cosmic microwave radiation 
spectrum together with the study of the large scale structure of the
universe. For a flat universe, WMAP combined with other astronomical 
data leads to \cite{Spergel:2006hy} $\sum_{i} m_i \leq 0.66 $ eV
($95\%$ CL). 

Neutrinoless double beta decay can also provide
information on the absolute scale  of the neutrino masses.
In the present framework, in the absence of additional lepton number
violating interactions, it provides a measurement of the
effective Majorana mass given by:
\begin{equation}
m_{ee} = \left| m_1 U_{e1}^2 +  m_2 U_{e2}^2  +  m_3 U_{e3}^2 \right|
\end{equation}
The present upper limit is $m_{ee} \leq 0.9$ eV
\cite{Fogli:2004as} from the Heidelberg-Moskow 
\cite{KlapdorKleingrothaus:2000sn} and the IGEX \cite{Aalseth:2002rf}
experiments. There is a claim of discovery of neutrinoless
double beta decay by the Heidelberg-Moscow collaboration
\cite{KlapdorKleingrothaus:2004wj}. Interpreted in terms of
a Majorana mass of the neutrino, this implies $m_{ee}$
between 0.12 eV to 0.90 eV. This result awaits confirmation
from other experiments and would constitute a major discovery.
It would set the scale of the neutrino masses and answer the still open 
question of whether or not neutrinos are Majorana particles.

Dirac type CP violation occurs whenever the imaginary 
parts of the quartets similar to those defined by Eq.~(\ref{qua}) 
for the quark sector:
\begin{equation}
I= \mbox{Im}\ U_{ij}U_{kl}U^*_{il}U^*_{kj}, \qquad 
(i \neq k, \ j \neq l)
\end{equation}
differ from zero. 
As previously emphasized, unitarity of $U$ insures that all imaginary parts 
are equal up to their sign. Therefore if any entry of the leptonic 
mixing matrix is zero, there is no Dirac-type CP violation in the
leptonic sector. On the other hand it can be easily 
verified from the structure of the indices of the quartets that
Majorana phases always cancel out in the quartets.
The simplest rephasing invariants 
in the leptonic sector include moduli of the matrix $U$ and quartets, 
as in the quark sector, and, in addition, products of the form
$ U_{ij} U^*_{ik}$, with no sum implied. Rephasing invariance of these
products results from the fact that the only rephasing transformations
allowed in this sector are $l_{L_i, R_i} \rightarrow 
e^{i \lambda_i} l_{L_i, R_i}$.
The minimal CP violating quantities are:
\begin{equation}
S_i \equiv    \mbox{Im} \ U_{ij} U^*_{ik} \qquad   
\mbox{no sum in i}
\end{equation}
provided the real part of $ U_{ij} U^*_{ik}$ is different from zero
\cite{AguilarSaavedra:2000vr}. 
Notice that the $S_i$ are sensitive to the presence of Majorana phases.
In the leptonic sector one can construct two types of unitarity triangles
\cite{AguilarSaavedra:2000vr}.The so-called Dirac triangles,
obtained through multiplication of rows of $U$, are similar to those 
in the quark sector. Majorana phases cancel in the product of each term and
under rephasing these triangles rotate in the complex plane as:
\begin{equation}
 \sum_{i} U_{ij} U^*_{kj} \rightarrow  e^{i(\lambda_i - \lambda_k)}
\sum_{i} U_{ij} U^*_{kj}
\end{equation}
Therefore their orientation has no physical meaning. They share a common area
proportional to $|I|$. The vanishing of this area does not imply that
the minimal CP violating quantities $S_i$ are zero and CP can still be 
violated. The second type of unitarity triangles are
constructed through multiplication of columns of $U$. 
These are the so-called Majorana triangles given by:
\begin{equation}
T_{jk} = U_{ej} U^*_{ek} + U_{\mu j} U^*_{\mu k} + U_{\tau j} U^*_{\tau k}
\end{equation}
In this case all terms in the sum are rephasing invariant. These 
triangles do not rotate under rephasing, and they are 
sensitive to the presence of Majorana type phases.
Figure 2 depicts an example of a hypothetical
Majorana triangle, obviously not based on current experimental 
observations. 
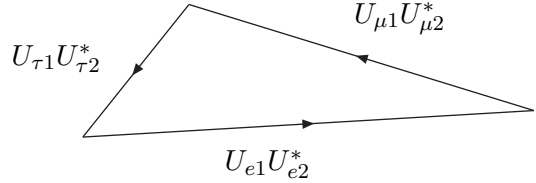
\begin{figure}[t]
\begin{center}
\begin{picture}(200,60)(0,0)
\ArrowLine(60,60)(20,10)
\Text(10,40)[]{$U_{\tau 1} U^*_{\tau 2}$}
\ArrowLine(20,10)(190,20)
\Text(90,0)[]{$U_{e1} U^*_{e2}$}
\ArrowLine(190,20)(60,60)
\Text(140,55)[]{$U_{\mu 1} U^*_{\mu 2}$}
\end{picture}
\end{center}
\caption{Majorana unitarity triangle $T_{12}$. Its orientation is fixed by the
Majorana phases and it cannot be rotated in the complex plane.}
\label{majtri}
\end{figure}

The Majorana triangles
provide the necessary and sufficient conditions for CP conservation
\cite{AguilarSaavedra:2000vr}: 

(i) Vanishing of their common area;

(ii) Orientation of all Majorana triangles along the 
direction of the real or of the imaginary axis.

The first condition implies that the Dirac phase vanishes. The 
second condition implies that the Majorana phases do not violate CP,
provided we are working with a real diagonal $d$ matrix, i.e., 
provided that the fields of the massive Majorana neutrinos satisfy 
self-conjugate relations which do not contain phase factors. 

CP conservation in the leptonic sector does not require that $U$
be a real matrix. In fact when a Majorana triangle is oriented
along the imaginary axis, Majorana phases are present but do not
violate CP. This is due to the existence of massive Majorana
neutrinos with opposite CP eigenvalues, also called CP parities
\cite{Wolfenstein:1981rk}. In order to illustrate this point
let us consider the following WB in which the charged leptons
have already been diagonalised:
\begin{equation}
{\cal L}_{mass} = - \nu^{0T}_{L} C \ m_{\nu} \ \nu^0_{L} -
\overline{l_L}\  d_l \  l_R + h. c. 
\end{equation}
It is obvious from previous analyses that CP is conserved
provided $m_\nu$ is real. Since Majorana mass terms are symmetric
by construction, the real matrix  $m_\nu$ can be diagonalised by an
orthogonal real transformation $O$ of the form:
\begin{equation}
O^T \ m_{\nu} \ O = \mbox{diag}\  (m_1, m_2, m_3)
\end{equation}
at this stage the $m_i$ are real but may be positive or negative.
Two cases are possible:

(i) all $m_i$ have equal sign;

(ii) one $m_i$ has a sign different from the other two.

Let us consider an example of case ii), for instance $m_2$ negative, 
$m_1$ and $m_3$ positive:
\begin{equation}
O^T \ m_{\nu} \ O = \mbox{diag}\  (\ |m_1|,\  -|m_2|,\  |m_3|\ )
\end{equation}
Positive masses are obtained by making the transformation:
\begin{equation}
K \ \mbox{diag}\  \left(\  |m_1|,\  -|m_2|,\  |m_3| \ \right) \ K = 
\mbox{diag}\ \left( \  |m_1|,\  |m_2|,\  |m_3|\  \right)
\end{equation}
with $K=$ diag (1, i, 1). All mass terms are now real positive
and diagonal and the mixing matrix $U$ is given by $U = O K$,
so that the charged current interaction can be explicitly
written as:
\begin{equation}
{\cal L}_W = - \frac{g}{\sqrt{2}} \left( 
\overline{e}, \  \overline{\mu} , \ \overline{\tau} \right)_L
\gamma^{\mu} \left( \begin{array}{ccc}  
 O_{11}  & i O_{12} & O_{13} \\
 O_{21}  & i O_{22} & O_{23} \\
 O_{31}  & i O_{32} & O_{33} \end{array} 
\right)
\left[ \begin{array}{c} 
\nu_1 \\ \nu_2 \\ \nu_3  \end{array} 
\right]_L \ W^+_{\mu}+h.c.
\end{equation}
A CP transformation of the mass eigenstates is of the form:
\begin{equation}
{\rm CP} \nu_{iL} ({\rm CP})^{\dagger} = {\eta_i}_{\ CP} \ C\  \nu^*_{iL},
\quad {\rm CP} l_{jL} ({\rm CP})^{\dagger} = {\eta^l_j}_{\ CP} \ C \  l^*_{jL}
\end{equation}
Since $d$ is a  Majorana mass matrix,
under CP  $d \rightarrow -d$, which  
requires that the $\eta_{CP} = \pm i$. The relative sign of 
$ {\eta_i}_{\ CP}$ is called the relative CP parity of the neutrinos.
The structure of the charged weak interactions fixes the relative
CP parities.  Suppose that all $O_{ij}$ are nonvanishing and take as
initial choice, e.g.:
\begin{equation}
{\rm CP} \nu_{1L} ({\rm CP})^{\dagger} = \ +i C  \nu^*_{1L}, 
\qquad \mbox{i.e.} \qquad {\eta_1}_{CP} = +i
\end{equation}
then $O_{11}$,  $O_{21}$,   $O_{31}$ force $\eta^e_{CP}$ =
 $\eta^\mu_{CP}$ =  $\eta^\tau_{CP}$ = +i \\
the couplings i $O_{j2}$ constrain $\nu_2$ to have ${\eta_2}_{CP} = -i$ \\
the couplings  $O_{j3}$ constrain $\nu_3$ to have ${\eta_3}_{CP} = +i$ \\

Therefore we are forced to assign a CP parity to $\nu_2$ which is
different from that of $\nu_1$ and $\nu_3$
It is clear from this discussion that CP parities can
only be defined in the CP conserving case. Furthermore only relative CP
parities have meaning.

It was shown in  \cite{Branco:1998bw} that, even in the limit of exact 
degeneracy and in the CP conserving case, leptonic mixing cannot be rotated
away provided neutrinos have different CP parities. \\

From the point of view of model building it is 
useful to derive WB invariant conditions 
for CP conservation in the 
leptonic sector analogous to those derived for the quark sector.
The procedure is analogous to the one outlined before and
was first applied to the leptonic sector in \cite{Branco:1986gr}.

Leptonic CP violation at low energies can be detected through
neutrino oscillations which are sensitive to the Dirac-type phase, 
but insensitive to the Majorana-type phases in the PMNS matrix. 
The strength of  Dirac-type CP violation can be obtained from the
following low energy WB invariant:
\begin{equation}
Tr[h_{eff}, h_l]^3= - 6i \Delta_{21} \Delta_{32} \Delta_{31}
{\rm Im} \{ (h_{eff})_{12}(h_{eff})_{23}(h_{eff})_{31} \} \label{trc}
\end{equation}
where $h_{eff}=m_{eff}{m_{eff}}^{\dagger} $, 
$ h_l = m_l m^\dagger_l $, and
$\Delta_{21}=({m_{\mu}}^2-{m_e}^2)$ with analogous expressions for
$\Delta_{31}$, $\Delta_{32}$. The righthand side of this equation
is the computation of this invariant in the special WB where the 
charged masses are real and diagonal. This invariant is analogous to 
the one presented in Eq.~(\ref{lll}), for the quark sector.
It can also be fully expressed in terms of physical observables since
\begin{equation}
{\rm Im} \{ (h_{eff})_{12}(h_{eff})_{23}(h_{eff})_{31} \} =
- \Delta m_{21}^2 \Delta m_{31}^2  \Delta m_{32}^2 I
\label{fjcp} 
\end{equation}
where $I$ is the imaginary part 
of an invariant quartet of the leptonic mixing matrix $U$
and is given by:
\begin{equation}
I \equiv {\rm Im}\left[\,U_{11} U_{22}
U_{12}^\ast U_{21}^\ast\,\right] 
= \frac{1}{8} \sin(2\,\theta_{12}) \sin(2\,\theta_{13}) \sin(2\,\theta_{23})
\cos(\theta_{13})\sin \delta\,, \label{Jgen1}
\end{equation}
A value for $ \theta_{13} $ close to the present experimental bound 
would be good news for the  prospects of detection of low energy leptonic 
CP violation, mediated through a Dirac-type phase and would correspond to
$I$ of order $10^{-2}$. Note that in the quark sector the 
corresponding $I$ is of the order $10^{-5}$.
Many other relations which are necessary conditions for CP invariance
can be derived. The Majorana character of the neutrinos provides
additional sources for CP violation. 
Selecting from the necessary conditions a subset of 
restrictions which are also sufficient for CP invariance
is in general not trivial. For three generations it
was shown that the following four conditions are sufficient
\cite{Branco:1986gr} to guarantee CP invariance:
\begin{eqnarray}
{\rm Im \ tr } \left[ h_l \; (m_{eff} \; m^*_{eff}) \;
( m_{eff} \; h^*_l \; m^*_{eff})\right] & = & 0 \label{41} \\
{\rm Im \ tr } \left[ h_l \; (m_{eff} \; m^*_{eff})^2 \;   
( m_{eff} \; h^*_l \; m^*_{eff}) \right] & = & 0 \label{42} \\
{\rm Im \ tr } \left[ h_l \; (m_{eff} \; m^*_{eff})^2 \   
( m_{eff} \; h^*_l \; m^*_{eff}) \; (m_{eff} \; m^*_{eff})\right]
 & = & 0 \label{43} \\
{\rm Im \ det } \left[ ( m^*_{eff} \; h_l \; m_{eff}) 
+ (h^*_l \;  m^*_{eff} \; m_{eff} )\right]   & = & 0 \label{44} 
\end{eqnarray}
provided that neutrino masses are nonzero and nondegenerate
\cite{special}. 
It can be easily seen that these conditions are trivially
satisfied in the case of complete degeneracy 
$(m_1  = m_2 = m_3 ) $. Yet there may still be CP violation
in this case, as stated before. In this limit 
 a necessary and sufficient condition 
\cite{Branco:1998bw} for CP invariance is:
\begin{equation}
G\equiv \ {\rm {Tr}}\left[ \ m^*_{eff}\cdot h_l\cdot m_{eff}\ ,\
h^*_l\right] ^3\ =\ 0.
\end{equation}

It is well known that the minimal structure that can lead to
CP violation in the leptonic sector is two generations of
lefthanded Majorana neutrinos  provided that 
their masses be non degenerate and that none of them
vanishes. In this case, it was proved \cite{Branco:1986gr} that
the condition
\begin{equation}
{\rm Im \ tr } \; Q = 0 \label{trq}
\end{equation}
with $Q = h_l  m_{eff}   m^*_{eff}  m_{eff} h^*_l  m^*_{eff}$
is a necessary and sufficient condition for CP invariance.

A more detailed discussion on WB invariant CP odd conditions relevant
for the leptonic sector and neutrino mass models can be found in
\cite{Branco:2004hu}.

\subsubsection{Leptogenesis}
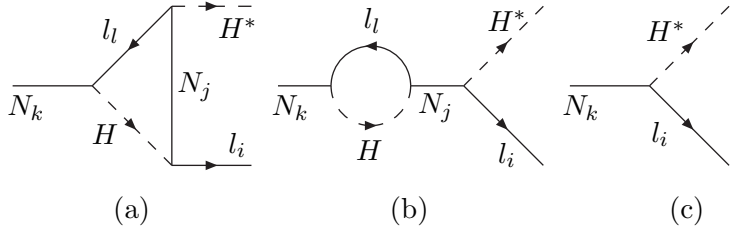
\begin{figure}[t]
\begin{center}
\begin{picture}(300,80)(0,0)
\Line(0,50)(30,50)
\DashArrowLine(60,80)(90,80){5}
\ArrowLine(60,20)(90,20)
\Line(60,80)(60,20)
\ArrowLine(60,80)(30,50)
\DashArrowLine(30,50)(60,20){5}
\Text(5,42)[]{$N_k$}
\Text(35,32)[]{$H$}
\Text(37,71)[]{$ l_l$}
\Text(69,50)[]{$N_j$}
\Text(85,72)[]{$H^\ast$}
\Text(85,28)[]{$ l_{i}$}
\Text(45,3)[]{(a)}
\Line(100,50)(120,50)
\DashArrowArc(135,50)(15,180,360){5}
\ArrowArc(135,50)(15,0,180)
\Line(150,50)(170,50)
\DashArrowLine(170,50)(200,80){5}
\ArrowLine(170,50)(200,20)
\Text(105,42)[]{$N_k$}
\Text(135,26)[]{$H$}
\Text(136,75)[]{$ l_l$} 
\Text(160,42)[]{$N_j$}
\Text(187,75)[]{$H^\ast$}
\Text(187,24)[]{$ l_{i}$}
\Text(150,3)[]{(b)}
\Line(210,50)(240,50)
\DashArrowLine(240,50)(270,80){5}
\ArrowLine(240,50)(270,20)
\Text(215,42)[]{$N_k$}
\Text(245,32)[]{$l_i$}
\Text(247,71)[]{$H^\ast$}  
\Text(255,3)[]{(c)}
\end{picture}
\end{center}
\caption{One-loop and tree diagrams contributing to the asymmetry
from the $N_k$ decay.}
\label{fig}
\end{figure}

The observed baryon asymmetry of the universe (BAU) is given by
\cite{Bennett:2003bz}:
\begin{equation}
\frac{n_{B}-n_{\overline B}}{n_{\gamma}}= (6.1 ^{+0.3}_{-0.2}) \times 10^{-10}.
\end{equation}   

There have been many attempts at explaining the origin of this 
asymmetry. Some reviews can be found in Ref.~\cite{BAU} 
One might wonder whether it could simply result from an 
initial condition with no need for further explanation. However, 
presently it seems very likely that our Universe went through a
period of inflation and inflation would have erased such a 
primordial asymmetry. Therefore this asymmetry must have been
generated after inflation. Sakharov \cite{Sakharov:1967dj}
conditions require that there be baryon number violation, C and CP 
violation and out of equilibrium dynamics. It is by now
established that the SM of electroweak interactions, where the 
Kobayashi- Maskawa mechanism is the only source of CP violation,
cannot produce a large enough asymmetry  \cite{Gavela:1994dt},
\cite{Huet:1994jb}, \cite{Anderson:1991zb}, \cite{Buchmuller:1993bq},
\cite{Kajantie:1995kf}, \cite{Fromme:2006cm}.
Furthermore a Higgs 
scalar mass above 80 Gev gives rise to a smooth phase transition and
therefore there is no out-of-equilibrium dynamics. The 
observed BAU requires the existence of physics beyond the SM. 
One of the most plausibe
explanations is Leptogenesis, since it relies on the only aspect of
physics beyond the SM that has already been observed, to wit neutrino
masses.

In this framework, the initial conditions are $B=0$ and $L=0$.
A CP asymmetry is generated through
out-of-equilibrium L-violating decays of heavy Majorana
neutrinos \cite{Fukugita:1986hr} leading to a lepton asymmetry
$L \neq 0$ while $B = 0$ is still maintained. 
Sphaleron processes \cite{Kuzmin:1985mm}, which are
$(B+L)$-violating and $(B-L)$-conserving, partially transform the 
the lepton asymmetry into a baryon asymmetry.
Figure 3 shows the tree level and one loop diagrams giving rise to
a lepton asymmetry, due to CP violation in the decay of the heavy
Majorana neutrinos. The lepton number asymmetry $\varepsilon _{N_{j}}$,
thus produced was computed by
several authors \cite{Liu:1993tg}, \cite{Flanz:1994yx}, \cite{Covi:1996wh},
\cite{Plumacher:1996kc}, \cite{Pilaftsis:1997jf}. 
Summing over all charged leptons
one obtains for the asymmetry produced by decay of the
heavy Majorana neutrino $N_j$ into the charged leptons 
$l_i^\pm$ ($i$ = e, $\mu$ , $\tau$):
\begin{small}
\begin{eqnarray}
\varepsilon _{N_{j}}
= \frac{g^2}{{M_W}^2} \sum_{k \ne j} \left[
{\rm Im} \left((m_D^\dagger m_D)_{jk} (m_D^\dagger m_D)_{jk} \right)
\frac{1}{16 \pi} \left(I(x_k)+ \frac{\sqrt{x_k}}{1-x_k} \right)
\right]
\frac{1}{(m_D^\dagger m_D)_{jj}}   \nonumber \\
= \frac{g^2}{{M_W}^2} \sum_{k \ne j} \left[ (M_k)^2
{\rm Im} \left((G^\dagger G)_{jk} (G^\dagger G)_{jk} \right)
\frac{1}{16 \pi} \left(I(x_k)+ \frac{\sqrt{x_k}}{1-x_k} \right)
\right]
\frac{1}{(G^\dagger G)_{jj}} \nonumber \\ 
\label{rmy}
\end{eqnarray}
\end{small}
where $M_k$ denote the heavy neutrino masses,
the variable $x_k$
is defined as  $x_k=\frac{{M_k}^2}{{M_j}^2}$ and
$ I(x_k)=\sqrt{x_k} \left(1+(1+x_k) \log(\frac{x_k}{1+x_k}) \right)$.
From Equation (\ref{rmy})
it can be seen that, when one sums over all 
charged leptons, the lepton-number
asymmetry is only sensitive to the CP-violating phases
appearing in $m_D^\dagger m_D$ in the WB, where $M_R $
is diagonal. Note that this combination is insensitive
to rotations of the lefthanded neutrinos. In many flavour models
the connection  between low energy CP violation and leptogenesis
is established in this basis, in scenarios where  $m_D$
verifies special constaints \cite{Frampton:2002qc}, \cite{Branco:2002kt},
\cite{Branco:2002xf},
\cite{Ibarra:2003up}, \cite{Branco:2005jr}.
In the general case it is not
possible to establish such a connection \cite{Branco:2001pq},
\cite{Rebelo:2002wj}.

Weak basis invariants relevant for leptogenesis  
were derived in \cite{Branco:2001pq}:
\begin{eqnarray}
I_1 \equiv {\rm Im Tr}[h_D H_R M^*_R h^*_D M_R]=0  \\
I_2 \equiv {\rm Im Tr}[h_D H^2_R M^*_R h^*_D M_R] = 0 \\
I_3 \equiv {\rm Im Tr}[h_D H^2_R M^*_R h^*_D M_R H_R] = 0 
\end{eqnarray}
with $h_D = m^\dagger_D m_D$ and  $H_R = M^\dagger_R M_R$
These constitute a set of necessary and sufficient conditions
in the case of three heavy neutrinos.  
Different expressions of the same type can be derived following
the same procedure.

The simplest leptogenesis scenario corresponds to
the case of heavy hierarchical neutrinos 
where $M_1$ is much smaller than $M_2$ and $M_3$.
In this limit only the asymmetry generated by the lightest
heavy neutrino is relevant, due to the existence of
washout processes, and  $\varepsilon _{N_{1}} $ can
be simplified  into:
\begin{equation}
\varepsilon _{N_{1}}\simeq -\frac{3}{16\,\pi v^{2}}\,\left( I_{12}\,
\frac{M_{1}}{M_{2}}+I_{13}\,\frac{M_{1}}{M_{3}}\right) \,,  \label{lepto3}
\end{equation}
where
\begin{equation}
I_{1i}\equiv \frac{\mathrm{Im}\left[ (m_D^{\dagger }m_D)_{1i}^{2}\right] }
{(m_D^{\dagger }\,m_D)_{11}}\ .  \label{lepto4}
\end{equation}
Thermal leptogenesis is a rather involved thermodynamical 
non-equilibrium process and depends on additional parameters.
In the hierarchical case the baryon asymmetry only depends on four
parameters \cite{Buchmuller:2002rq}, \cite{Buchmuller:2002jk}, 
\cite{Davidson:2003fj}, \cite{Davidson:2002qv}:
the mass $M_{1}$ of the lightest heavy
neutrino, together with the corresponding 
CP asymmetry $\varepsilon _{N_{1}}$ in
their decays, as well as the effective neutrino mass $\widetilde{m_{1}}$  
defined as
\begin{equation}
\widetilde{m_{1}}=(m_D^{\dagger }m_D)_{11}/M_{1}  \label{mtil}
\end{equation}
in the weak basis where $M_R$ is diagonal, real and positive. Finally, 
the baryon asymmetry depends also on the 
sum of all light neutrino masses squared, 
${\overline{m}}^{2}=m_{1}^{2}+m_{2}^{2}+m_{3}^{2}$, since it has been 
shown that this sum controls an important class of washout processes.

Leptogenesis is a non-equilibrium process
that takes place at temperatures $T\sim M_{1}$.
This imposes an upper bound on the effective neutrino mass 
$\widetilde{m_{1}}$ given by the ``equilibrium neutrino mass''
\cite{Kolb:1990vq}, \cite{Fischler:1990gn}, \cite{Buchmuller:1992qc}:
\begin{equation} 
m_{*}=\frac{16\pi ^{5/2}}{3\sqrt{5}}g_{*}^{1/2}\frac{v^{2}}{M_{Pl}}\simeq
10^{-3}\ \mbox{eV}\; ,  \label{enm}
\end{equation}  
where $M_{Pl}$ is the Planck mass ($M_{Pl}=1.2\times 10^{19}$ GeV),
$v=\langle \phi ^{0}\rangle /\sqrt{2}\simeq 174\,$GeV is the weak scale 
and $g_{*}$ is the effective number of relativistic degrees of
freedom in the plasma and equals 106.75 in the SM case.
Yet, it has been shown \cite{Buchmuller:2003gz}, \cite{Buchmuller:2004nz},
\cite{Buchmuller:2004tu}, \cite{Giudice:2003jh} that successful leptogenesis
is possible for $\widetilde{m_{1}} < m_{*}$ as well as
$\widetilde{m_{1}} > m_{*}$, in the range from 
$\sqrt {\Delta m^2_{12}}$ to  $\sqrt {\Delta m^2_{23}}$. 
The square root of the sum
of all neutrino masses squared ${\overline{m}}$ is constrained, in the
case of normal hierarchy, to be below 0.20 eV  \cite{Buchmuller:2003gz}, 
\cite{Buchmuller:2004nz}, \cite{Buchmuller:2004tu}, 
which corresponds to an upper bound on light neutrino masses
very close to 0.10 eV. This result is sensitive to radiative
corrections which depend on top and Higgs masses as well as on
the treatment of thermal corrections. 
In \cite{Giudice:2003jh} a slightly higher value of 0.15 eV is found.
From Eq.~(\ref{lepto3}) a lower bound 
on the lightest heavy neutrino mass $M_{1}$
is derived. Depending on the cosmological scenario, the range for
minimal  $M_{1}$ varies from order $10^7$ Gev to $10^9$ Gev 
\cite{Buchmuller:2002rq}, \cite{Giudice:2003jh}. 

It was pointed out recently \cite{Barbieri:1999ma}, \cite{Endoh:2003mz}
\cite{Fujihara:2005pv}, \cite{Pilaftsis:2005rv}, \cite{Vives:2005ra}
\cite{Abada:2006fw}, \cite{Nardi:2006fx}, \cite{Abada:2006ea},  
\cite{Blanchet:2006be}
that there are cases where flavour matters and the commonly
used expressions for the lepton asymmetry, which depend on
the total CP asymmetry and one single efficiency factor, may fail
to reproduce the correct lepton asymmetry. In these cases, the calculation
of the baryon asymmetry produced by thermal leptogenesis
with hierarchical righthanded neutrinos must take into consideration
flavour dependent washout processes. 
As a result, in this case,
the previous upper limit on the light neutrino masses does
not survive and leptogenesis can be made  viable with neutrino masses
reaching the cosmological bound of $\sum_{i} m_i \leq 0.66 $ eV.
The lower bound on $M_1$ does not move much with the inclusion of 
flavour effects.
The separate lepton $i$ family asymmetry generated from the decay 
of the $k$th heavy Majorana neutrino  depends on the combination 
\cite{Fujihara:2005pv}
Im$\left( (m_D^\dagger m_D)_{k k^\prime}(m_D^*)_{ik} (m_D)_{ik^\prime}\right) $
as well as on 
Im$\left( (m_D^\dagger m_D)_{k^\prime k}(m_D^*)_{ik} (m_D)_{ik^\prime}\right) $
summing over all leptonic flavours $i$ the second term becomes
real so that its imaginary part vanishes and the first term 
gives rise to the combination 
Im$\left((m_D^\dagger m_D)_{jk} (m_D^\dagger m_D)_{jk} \right)$
that appears in Equation (\ref{rmy}). Flavour effects bring new sources 
of CP violation to leptogenesis and the possibility of having a 
common origin for CP violation at low energies and 
for leptogenesis \cite{Pascoli:2006ie}, 
\cite{Branco:2006ce}, \cite{Branco:2006hz}, \cite{Uhlig:2006xf}.

We have just refered to the minimal scenario for thermal
leptogenesis. For a review including other scenarios 
see \cite{Buchmuller:2005eh}.
The case of resonant leptogenesis is a
remarkable alternative allowing for much lighter
heavy neutrinos, and has recently raised a considerable interest 
\cite{Pilaftsis:1997jf}, \cite{Pilaftsis:2003gt}, \cite{Pilaftsis:2005rv}.
One elegant way of obtaining the required smallness of the mass splitting 
of the heavy neutrinos is through radiative effects induced by 
renormalization group running  \cite{GonzalezFelipe:2003fi}, 
\cite{Turzynski:2004xy}, \cite{Branco:2005ye}, \cite{Branco:2006hz}.

There are many other interesting scenarios 
for leptogenesis which we do not cover here.

\section*{Acknowledgements}
G. C. B. thanks the Organizers of the 47th Cracow School of Theoretical 
Physics, which took place in Zakopane, Poland for the warm hospitality
and the stimulating scientifical environment provided both by the
talks and by the informal discussions.
This work was partially supported by Funda\c c\~ ao para a
Ci\^ encia e a  Tecnologia (FCT, Portugal) through the projects
POCTI/FNU/44409/2002, PDCT/FP/63914/2005, PDCT/FP/ \\  63912/2005 and
CFTP-FCT UNIT 777 which are partially funded through POCTI 
(FEDER). M. N. R. is presently at CERN on sabbatical leave.
G.C.B. and M.N.R. are grateful for the warm hospitality of
the CERN Physics Department (PH) Theory (TH) where this work
was finalised.

\end{document}